\documentclass[prd,preprint,onecolumn,superscriptaddress,nofootinbib,amsfonts]{revtex4-2}

\usepackage[utf8]{inputenc}
\usepackage{amsmath}
\usepackage{amsfonts}
\usepackage{verbatim}
\usepackage{amssymb}
\usepackage{graphicx}
\usepackage{slashed}
\usepackage{epsfig}
\usepackage{epstopdf}
\usepackage{adjustbox}
\newcommand{\ii}{\mathrm{i}}

\newcommand{\beq}{\begin{eqnarray}}
\newcommand{\eeq}{\end{eqnarray}}
\newcommand{\tr}{\mbox{tr}}
\def\g{\gamma}
\def\a{\alpha}

\def\m{\mu}
\def\n{\nu}
\def\gm{\gamma^{\mu}}
\def\gn{\gamma^{\nu}}
\def\ga{\gamma^{\alpha}}

\def\arctanh{\mbox{arctanh}}
\def\emna{\varepsilon^{\mu\nu\alpha}}

\def\eq{&=&}
\def\sgn{\mbox{sgn}}

\def\nn{\nonumber}
\def\sp{\slashed{p}}
\def\sk{\slashed{k}}
\usepackage{xcolor} \definecolor{darkgreen}{rgb}{0,.5,0}
\usepackage[colorlinks,filecolor=blue,citecolor=darkgreen,unicode]{hyperref}
\begin{document}
\begin{titlepage}









\title{Parity anomaly with impurities and the Pauli--Villars subtraction}
\author{Ozório Holanda}\email{netoholanda91@gmail.com}
\affiliation{Centro de Matemática, Computação e Cognição - Universidade Federal do ABC, Santo André, SP, Brazil}
\author{ Ren\'e Meyer}\email{rene.meyer@uni-wuerzburg.de}
\affiliation{Institute for Theoretical Physics and Astrophysics, Julius--Maximilians--Universität Würzburg, Am Hubland, 97074 Würzburg, Germany}
\affiliation{Würzburg-Dresden Cluster of Excellence ct.qmat}
\author{Dmitri Vassilevich}\email{dvassil@gmail.com}
\affiliation{Centro de Matemática, Computação e Cognição - Universidade Federal do ABC, Santo André, SP, Brazil}

\begin{abstract}
We calculate the anomalous part of the polarization tensor of Dirac fermions in $2+1$ dimensions in the presence of impurities described by the scattering rate $\Gamma$ for arbitrary external frequency and momenta. We consider two different versions of the Pauli--Villars subtractions and discuss their physical consequences. 
\end{abstract}

\maketitle

\end{titlepage}

\section{Introduction}

Due to the parity anomaly \cite{Redlich:1983dv,Niemi:1983rq} (see also \cite{Deser:1981wh}), the one-loop effective action for Dirac fermions in $2+1$ dimensions contains a parity violating Chern--Simons part. From phenomenological point of view, this means the appearance of a quantum anomalous Hall effect, i.e. of a Hall type conductivity in the absence of an external magnetic field. The momentum dependence of the anomalous part of the polarization tensor was analyzed in \cite{Appelquist:1986qw}. These papers were followed by a very interesting development in quantum field theory. To learn about general aspects of the Chern--Simons theory the reader may consult \cite{Dunne:1998qy}.

An intriguing feature of the parity anomaly is that it leads to a Chern--Simons term with the weight $\pm 1/2$ (or, equally, to the Hall conductivity being $\pm 1/2$ of the Hall quantum). Such a term is not invariant under large gauge transformations which has been a source of confusion for a long period. This apparent contradiction was finally resolved in \cite{Deser:1997gp,Deser:1997nv} where it was demonstrated that the Ward identities corresponding to large gauge transformations contain nonperturbative contributions capable to restore gauge invariance of the effective action.

Theoretical investigations of quantum anomalies are highly relevant in condensed matter systems, see \cite{Dudal:2021ret}. For the benefit of the typical high energy physics reader, we here given an overview over the main developments in this direction: Quantum anomalies are tied to relativistic fermions, which arise in topologically protected approximately relativistic band crossings such as topological insulators with 2+1-dimensional Dirac surface states \cite{KaneMele2005,BHZ2006,Molenkamp2007,FuKaneMele2007,LiuHughesQi2008,Roy2009,hsieh2008topological,xia2009observation,Chenetal2009,KnezDuSullivan2011}, or Weyl and Dirac \cite{WanAriVishwanathSavrasov2011,XuWengWangDaiFang2011,Borisenkoetal2014,liu2014stable,LiuChenetal2014,neupane2014observation} semimetals with 3+1-dimensional Weyl fermions as band crossings. In particular, the parity anomaly directly contributes to the quantum anomalous Hall effect, a Hall effect in the absence of external magnetic fields, in two- dimensional quantum anomalous Hall (QAH) insulators such as (Hg,Mn)Te quantum wells \cite{LiuZhang2008,QiZhang2008} or magnetically doped (Bi,Sb)Te thin films \cite{YuetalZhong2010,ChangetalXue2013}. The DC quantum anomalous Hall conductivity has a quantized part that is directly induced by the parity anomaly, i.e. the coefficient of the Chern-Simons term in the low energy effective action. In real-world condensed matter systems such as QAH insulators, the band structure can contain UV relevant terms \cite{BHZ2006} which serve as a UV regulator and contribute to the quantized as well as non-quantized parts of the quantum anomalous Hall conductivity \cite{tutschku2020momentum}. They preserve large gauge invariance and break parity, hence contributing as a UV regulator to the parity anomaly \cite{tutschku2020momentum}. From the point of view of the band structure, the quantized part of the anomalous Hall conductivity can be related to the presence of momentum space Berry curvature in the wave functions at the Dirac band touching points \cite{bernevig2013topological}, or equivalently to a quantized winding number in the Brillouin zone \cite{thouless1982quantized}. There are other non-quantized,  $(T,\mu)$ dependent contributions to the anomalous Hall conductivity as well \cite{tutschku2020temperature}.  Experimental signatures of the parity anmomaly in 2+1-dimensional Dirac materials have been discussed in e.g. \cite{PhysRevLett.123.226602,hu2022signature}. 

In condensed matter physics, there are several ways to describe impurities depending on the properties of a particular material \cite{TkachovBook}. The usual assumption is that impurities provide a randomly distributed non-periodic short-range scattering potential, on which the conduction electrons in a solid scatter in an energy and charge conserving way, but loose momentum on these impurities. For small impurity densities, the momentum-relaxing effect of impurity scattering, together with the details of the potential, can be summarized in a momentum relaxation term on the right hand side of the translation ward identity. An analogous relaxation time approximation is possible for weak breaking of translational symmetry in AdS/CFT models of strongly correlated quantum matter, for a recent review c.f.~\cite{zaanen2015holographic,hartnoll2018holographic}. In a particular limit, the AdS/CFT correspondence \cite{Maldacena1997,Gubser1998,Witten1998anti} is a duality between strongly interacting quantum systems on the one side, and semiclassical gravity in one additional dimension on the other side (for a review of the correspondence, c.f.~e.g.~\cite{ammon2015gauge}). In AdS/CFT, breaking of translational symmetry can be implemented in several ways \cite{Horowitz:2012gs,Vegh:2013sk,Davison:2013jba,Blake:2013bqa,Andrade:2013gsa,Gouteraux:2014hca,Donos:2012js,Donos:2012js,Erdmenger:2015qqa}. For small induced momentum relaxation rate these models are all equivalent to a theory with a massive graviton \cite{Blake:2013owa}, which is responsible for the relaxation term on the right hand sinde of the translational Ward identity. 

Long-range charged impurities or long-range disorder potentials such as the ones generated by the charge puddles in graphene need a different treatment \cite{TkachovBook},  by introducing a scattering rate $\Gamma$ which enters the propagator of quasiparticles through substitution of the temporal momentum $p_0\to p_0+\ii \Gamma \sgn (p_0)$. This approach to the impurities was used in calculations of the anti-symmetric part of polarization tensor in graphene in the presence of external magnetic field in e.g. \cite{Gusynin:2005iv,Gusynin:2006ym,Gusynin:2006gn,Fialkovsky:2012ee}. The computations of \cite{Fialkovsky:2012ee} are in a very good agreement with the measurements \cite{Crassee_2010} of giant Faraday rotation in graphene. Note that in these calculations an antisymmetric part of the polarization tensor was caused by the presence of an external magnetic field. The anomalous Hall conductivity of graphene cancels between generations of fermions. This explains why the anomalous Hall contributions with impurirites described by scattering rate $\Gamma$ were neglected at that time. However, new applications including Hall conductivities of surface states of topological materials put the problem forward again.

The purpose of this work is to close an important gap in the literature by computing the anomalous part of polarization tensor for fermions in $2+1$ dimensions in the presence of impurities described by the scattering rate $\Gamma$. We do not relate this computation to any particular material. Our main attention is on the Quantum Field Theory aspects. In particular, we study the Pauli--Villars subtraction scheme and suggest two versions of this scheme leading to qualitatively different dependence of the parity anomaly on $\Gamma$. The first scheme, consists in subtracting the contribution of a regulator field with mass $M$ with taking subsequently the limit $|M|\to\infty$. In the second scheme, we also take the limit $\Gamma_R\to\infty$, where $\Gamma_R$ is the impurity parameter for the regulator field, keeping the ratio $\Gamma_R/M$ fixed. From the Quantum Field Theory perspective, the difference between two schemes is merely a renormalization ambiguity. We perform the computations for arbitrary values of external frequency and momenta which is not common for current literature although this is important for evaluation of the Casimir force, for example. 

This paper is organized as follows. In the next section, we compute the unregularaized anomalous polarization tensor. Two possible PV subtraction schemes are analyzed in Section \ref{sec:PV}. The results of this work are briefly discussed in Section \ref{sec:dis}.

\section{Polarization tensor in presence of impurities}\label{sec;Pol}

We start with the action
\beq
S = \int d^3 x \Bar{\Psi}\slashed{D}\Psi,
\eeq
where $\slashed{D} = \ii\Tilde{\gamma}^{\m}(\partial_\mu+ie A_{\mu}) -m$. A twiddle over a 3-vector means that that the spatial components are rescaled with the Fermi velocity. In particular, 
\beq
\Tilde{\gamma}^0 = \gamma^{0} , \,\,\,\, \Tilde{\gamma}^{i} = v_F \gamma^{i}
\eeq
The Greek letters label the spacetime coordinates, $\mu = 0,1,2$, while the Latin letters denote the spatial coordinates, $i,j=1,2$. We use the metric $g^{\m\n}= \mbox{diag}(1,-1,-1)$. We fix $\tr[\gm\gn\ga]= 2\ii\emna$. Here, $\emna$ is the Levi-Civita tensor, and $\ii:=\sqrt{-1}$.

The one-loop effective action due to quantum fermions in the second order of electromagnetic field $A$ reads
\begin{equation}
    S_{\mathrm{eff}}=\frac 12 \int \frac {d^3k}{(2\pi)^3} A_\mu(-k)\Pi^{\mu\nu}(k)A_\nu(k),\label{Seff}
\end{equation}
where the polarization tensor is given by
\begin{equation}
 \Pi^{\mu\nu}(k,m) = \ii e^2\int\frac{d^3p}{(2\pi)^3} \tr[\tilde\g^\mu G_f(p,m)\tilde\g^\nu G_f(p+k,m)]   
\end{equation}
with the fermion propagator
\beq
G_f(p, m) = \frac{1}{\slashed{\tilde p}-m} = \frac{(\slashed{\tilde p}+m)}{\tilde p^2 - m^2},\label{eq3}
\eeq
where $\slashed{\tilde p}:=\tilde\gamma^\mu p_\mu$. In the presence of impurities, the temporal momentum has to be replaced as
\begin{equation}
    p_0\to \hat p_0=p_0+\ii \Gamma\, \mathrm{sgn}\, p_0.\label{phat}
\end{equation}
Here $\Gamma$ is an impurity scattering rate, $\Gamma>0$.

Note, that the replacement (\ref{phat}) done in the propagator alone breaks gauge invariance. Indeed, such propagator can be formally obtained from a Dirac action containing an additional term with $\mathrm{sgn}(-\ii \partial_0)$. To ensure gauge invariance, any derivative has to be accompanied by an electromagnetic potential. Thus, one expects something like $\mathrm{sgn}(-\ii (\partial_0+eA_0))$. Giving a precise meaning to such a term is not easy, but obviously it should modify the couplings of fermions to $A_0$ and lead to new vertices in the Feynman rules involving $A_0$. To avoid complications, we will consider only the diagrams with external legs corresponding to the spatial components $A_j$, i.e the $\Pi^{ij}$ part of the polarization tensor. This will be enough for our purposes since the full polarization tensor can be recovered through the transversality condition. Thus, we will evaluate
\begin{equation}
    \Pi^{ij}(k)=  \ii e^2\int\frac{d^3p}{(2\pi)^3} \tr[\tilde\g^i G_f(\widehat{p},m)\tilde\g^j G_f(\widehat{p+k},m)] ,
\end{equation}\label{Piij}
where $\widehat{p}^{\, \mu} := (p_0 + \ii \Gamma\, \sgn{(p_0)}, \Vec{p})$.

To analyze the dependence of polarization tensor on the Fermi velocity one can use the following simple trick \cite{Fialkovsky:2009wm}. Let us change the integration variable $p\to \tilde{p}$ in (\ref{Piij}). The Jacobean factor $v_F^{-2}$ is canceled the factors of $v_F$ in $\tilde\g^i$ adn $\tilde \g^j$, so that one obtains
\begin{equation}
    \Pi^{ij}(v_F;k)=\Pi^{ij}(v_F=1,\tilde k) \label{PivF}
\end{equation}
Thus, from now on we set
\begin{equation}
    v_F=1.
\end{equation}

Thus we have
\beq
\Pi^{ij}(k,m,\Gamma) 
 &=& \ii e^2\int\frac{d^3p}{(2\pi)^3}\frac{ \tr[\g^i (\widehat{\slashed{p}}+m)\g^j (\widehat{\sp+\sk}+m)]}{(\widehat{p}^2-m^2)((\widehat{p+k})^2-m^2)}.\label{Piij1}
\eeq

We are interested in the anomalous part $\Pi_{\mathrm{odd}}$ of the polarization tensor which changes the sign under the change of orientation of spacetime. This is the part which is antisymmetric in the indices $i,j$ and proportial to $\varepsilon^{ij}\equiv \varepsilon^{ij0}$. After computing the traces over spinor indices and selecting relevant tensor structures we obtain
\beq
    \Pi_{\mathrm{odd}}^{ij}(k,m,\Gamma)=\varepsilon^{ij}k_0\left[ \zeta(k,m,\Gamma) +  \chi(k,m,\Gamma)\right],
\eeq
where
\beq
\zeta(k,m,\Gamma)&=&-2me^2\int\frac{d^3p}{(2\pi)^3}\frac{1 }{((p_0 + \ii \Gamma\, \sgn{(p_0)})^{\,2}-\Vec{p}^{\,2}-m^2)}\nonumber\\
 &\times&\frac{1}{((p_0 + k_0 + \ii \Gamma\, \sgn{(p_0+k_0)})^{2}-(\Vec{p}+\Vec{k})^2-m^2)}
\eeq
and
\beq
\chi(k,m,\Gamma)&=&-2\ii me^2{\frac \Gamma {k_0}} \int\frac{d^3p}{(2\pi)^3}\frac{\sgn{(p_0 +k_0)}-\sgn{(p_0)}}{((p_0 + \ii \Gamma\, \sgn{(p_0)})^{\,2}-\Vec{p}^{\,2}-m^2)}\nonumber\\
&& \times \frac {1}{ (p_0 + k_0 + \ii \Gamma\, \sgn{ (p_0+k_0)} )^2-(\Vec{p}+\Vec{k})^2-m^2 } \,. 
\eeq
We use the Feynman formula
\beq
\frac{1}{AB}=\int_{0}^{1} dx {1\over(Ax +(1-x)B)^2} .
\eeq
Quite remarkably, the expression under integral above has no poles in the whole integration region despite the presence $\Gamma$. We thus have
\beq
 \zeta(k,m,\Gamma)=-2me^2\int_{0}^{1} dx \int \frac{dp_0}{2\pi}\int\frac{d^2 \Vec{l}}{(2\pi)^2}\frac{1}{(\Vec{l}^2-M)^2}
 \eeq
and
\beq
\chi(k,m,\Gamma)=- 2i me^2{\Gamma\over k_0} \int_{0}^{1} dx \int \frac{dp_0}{2\pi}(\sgn(p_0 +k_0)-\sgn(p_0))\int \frac{d^2 \Vec{l}}{(2\pi)^2}\frac{1}{(\Vec{l}^2-M)^2},
\eeq
with $\Vec{l}=\Vec{p}+\Vec{k}x$ and $M =  (p_0 + \ii \Gamma\, \sgn{(p_0)})^2(1-x)+x(p_0+k_0+ \ii \Gamma \sgn(p_0 + k_0))^2 -\Vec{k}^2 x (1-x)-m^2.$

After performing the $\Vec{l}$ integration we obtain
\begin{eqnarray}
&&\zeta(k,m,\Gamma)=\frac{me^2}{4\pi^2}\, \int_{0}^{1} dx \int dp_0\frac{1}{M},\label{ze33}\\
&&\chi(k,m,\Gamma)=\frac{\ii me^2\Gamma}{4\pi^2k_0}\, \int_{0}^{1} dx\int dp_0\frac{\sgn{(p_0 +k_0)}-\sgn(p_0)}{ M} .\label{chi33}
\end{eqnarray}
In the presence of $\Gamma$, the $p_0$ integral cannot be done by computing the residues and requires somewhat more attention. In the expression (\ref{ze33}), one has to divide the integration region in three intervals depending on the signs of $p_0$ and $p_0+k_0$. In (\ref{chi33}) only one of such intervals contribute. After performing the integrations, one obtains
\beq
\zeta(k,m,\Gamma)&=& \frac{1}{4\pi^2}me^2\int_{0}^{1} dx\left[ \frac{\pi-2 \,\mbox{arctan}\left(\frac{k_0 x+\ii\Gamma }{\sqrt{-m^2 +k^2 x (1-x) }}\right)
}{ \sqrt{-m^2 +k^2 x (1-x) }}\right.\nonumber\\
&+&\left. \frac{2\,\mbox{arctan}\left(\frac{(k_0 + 2\ii \Gamma) x-i\Gamma}{\sqrt{-m^2 +((k_0 + 2\ii\Gamma)^{2}-\Vec{k}^2) x(1-x)}}\right)}{\sqrt{-m^2 +((k_0 + 2\ii\Gamma)^{2}-\Vec{k}^2) x(1-x)}}\right]\label{eq57}
\eeq
and
\beq
\chi(k,m,\Gamma)=\frac{\ii}{4\pi^2}me^2 \frac{\Gamma}{k_0} \int_{0}^{1} dx \frac{4\,\mbox{arctan}\left(\frac{(k_0 + 2\ii \Gamma) x-\ii\Gamma}{\sqrt{-m^2 +((k_0 + 2\ii\Gamma)^{2}-\Vec{k}^2) x(1-x)}}\right)}{\sqrt{-m^2 +((k_0 + 2\ii\Gamma)^{2}-\Vec{k}^2) x(1-x)}}.
\eeq

\subsection{Limits of the polarization tensor}\label{sec:lim}
Before going on with the renormalization, let us consider some limits of unrenormalized polarization tensor. In the $m\to 0$ limit with all other parameters staying finite one immediately obtains
\begin{equation}
    \Pi^{ij}_{\mathrm{odd}}(k,0,\Gamma)=0.\label{mto0}
\end{equation}
Another important limit is $\Gamma\to 0$,
\begin{equation}
   \Pi_{\mathrm{odd}}^{ij}(k,m,0)= -\frac{\ii e^2}{4\pi}\varepsilon^{ij} k_{0} \frac{2m}{|k|} \arctanh\left(\frac{|k|}{2|m|}\right),
   \label{Gammato0}
\end{equation}
which reproduces the known result \cite{Appelquist:1986qw}. Do not confuse $\arctanh$ in this formula with $\arctan$ in Eq.\ (\ref{eq57}).

In the limit when both $m$ and $\Gamma$ become large but their fraction is kept constant, we have
\beq
\lim_{\lambda\to\infty}{\Pi}_{\mathrm{odd}}^{ij}(k,\lambda m,\lambda\Gamma) &=&\frac{1}{4\pi^2}me^2\varepsilon^{ij}k_{0}\left(\frac{-\ii\pi +2\ii\, \mbox{arctan}\left(\frac{\Gamma}{\vert m\vert}\right)}{\vert m \vert}\right)\nonumber\\
&-&\frac{1}{4\pi^2}\ii\Gamma me^2\varepsilon^{ij}k_0\left(\frac{2}{\Gamma^2 + m^2}\right).\label{mGamma}
\eeq
(This formula is easier to obtain from (\ref{ze33}) and (\ref{chi33}) where the integration over $p_0$ has not been done yet.) 

The same formula (\ref{mGamma}) describes the limit when both $k_0$ and $|\Vec{k}|$ are small as compared to $m$ and $\Gamma$. This formula also allows to obtain the limits
\begin{equation}
    \lim_{|m|\to\infty}{\Pi}_{\mathrm{odd}}^{ij}(k,m,\Gamma)=-\frac{\ii e^2}{4\pi}\, \frac{m}{|m|}\varepsilon^{ij}k_{0} \label{mtoinfty}
\end{equation}
and
\begin{equation}
  \lim_{\Gamma\to\infty}{\Pi}_{\mathrm{odd}}^{ij}(k,m,\Gamma)  =0.\label{Gammatoinfty}
\end{equation}

To recover the dependence of polarization tensor on Fermi velocity it is sufficient to replace $k\to \tilde k=(k_0,v_F\Vec{k})$ in the formulas given above. The full anomalous polarization tensor may be obtained by solving the conservation condition $k_\mu \Pi_{\mathrm{odd}}^{\mu\nu}=0$. One obtains $\Pi_{\mathrm{odd}}^{0j}=-(k_i/k_0) \Pi_{\mathrm{odd}}^{ij}$ and $\Pi_{\mathrm{odd}}^{00}=0$.

\section{Pauli--Villars subtractions}\label{sec:PV}
As has been noted already by Redlich \cite{Redlich:1983dv}, although the parity-odd part of polarization tensor is non-divergent, one has to apply a Pauli--Villars subtraction to get a correct result. The usual prescription consists of subtracting from the non-renormalized polarization tensor a contribution from a fermion field have exactly the same parameters except for the mass $M$ and taking the limit $|M|\to\infty$ at the end of calculations. We additionally assume that $M$ has the same sign as $m$ so that to pass from $m$ to $M$ one does not need to cross the gapless phase. This assumption is not essential. The opposite limit can be analysed along the same lines. We will call this scheme
PV1 (since there also be a PV2). Basically, this prescription boils down to subtracting (\ref{mtoinfty}) from the polarization tensor.

For further discussion, is it convenient to introduce a quantity $\sigma$,
\begin{eqnarray}
 \Re\left[  \frac{\Pi^{ij}_{\mathrm{odd}}}{\ii k_0}\right]=\varepsilon^{ij}\, \frac{e^2}{2\pi}\, \sigma
\end{eqnarray}
which is nothing else than the anti-symmetric (Hall) conductivity measured in the units of Hall quantum $e^2/(2\pi)=e^2/h$. The imaginary part of conductivity is not affected by the Pauli--Villars subtraction and thus will not be considered here.

We immediately obtain the zero-gap result
\begin{equation}
    \sigma_{\mathrm{PV1}}(k,0,\Gamma)=\frac 12 \, \frac{m}{|m|}\,,\label{PV1m0}
\end{equation}
which comes exclusively from the PV regulator field\footnote{The sign on the right hand side is a consequence of our assumption $\sgn\, M=\sgn \, m$. In general case, the sign factor in (\ref{PV1m0}) is given by $\sgn\, M$.},  does not depend on $\Gamma$, and represents the classical value of parity anomaly. Also, the limit $\Gamma\to 0$ reproduces a known result,
\begin{equation}
    \sigma_{\mathrm{PV1}}(k,m,0)=\frac 12 \left[  \frac{m}{|m|} -\frac{2m}{|k|} \arctanh\left(\frac{|k|}{2|m|}\right)\right] \,.\label{PV1Gamma0}
\end{equation}
We also have
\begin{equation}
  \lim_{\Gamma\to\infty}\sigma_{\mathrm{PV1}}(k,m,\Gamma)  = \frac 12 \, \frac{m}{|m|}\,.\label{PV1Gammatoinfty}
\end{equation}

In the scheme PV1, the mass of regulator field goes to infinity while the parameter $\Gamma$ which describes impurities remains fixed. Thus, relative to the mass, the impurities become negligible. Since $\Gamma$ is a phenomenological parameter, one can also consider other prescriptions for the behavior of the impurity parameter $\Gamma_R$ for the regulator field. A reasonable choice seems to be to fix the ratio $\Gamma_R/M$ while taking the limit $\Gamma_R,\ M\to\infty$. In other words, we take the limit $m\to\infty$ keeping all \emph{dimensionless} parameters like $e$ and $\Gamma_R/M$ fixed. We call this scheme PV2. In this scheme, we need to subtract the expression (\ref{mGamma}) from the unrenormalized polarization tensor. Let us discuss physical consequences of the PV2 subtraction.

\begin{figure}[h]
    \centering
    \includegraphics[width=16cm]{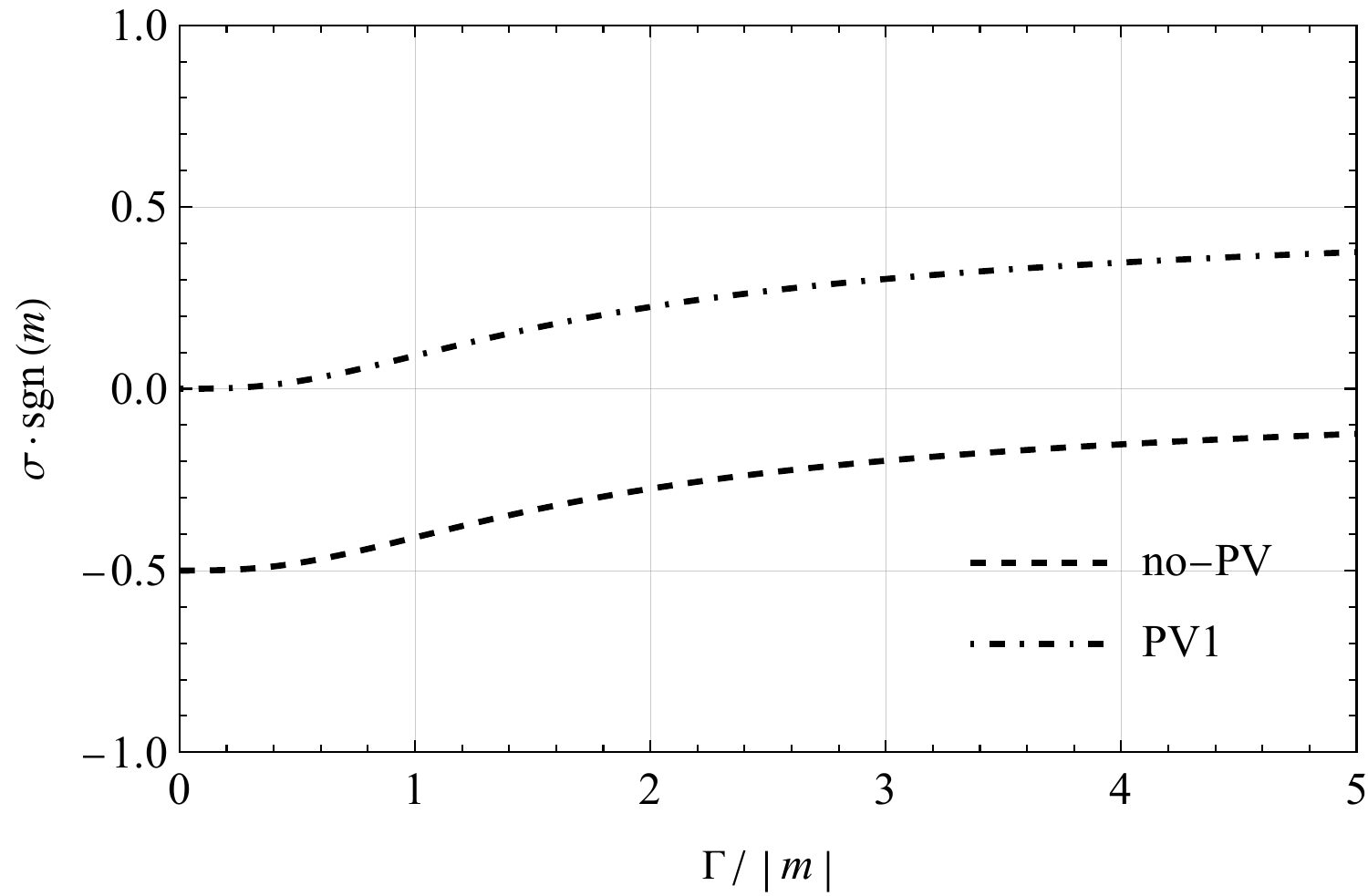}
    \caption{The anomalous conductivity $\sigma$ at $\Vec{k}=0$, $k_0=0$ as a function of $\Gamma/|m|$ without subtraction (dashed line) and with PV1 subtraction (dashed-dotted line). Since $\sigma_{\mathrm{PV2}}(0,m,\Gamma)=0$ the corresponding plot is not included. }
    \label{fig:k00}
\end{figure}

If $m\neq 0$, there are simple analytic formulas for $k=0$ which we present here for the sake of completeness.
\begin{equation}
    \sigma(0,m,\Gamma)=\frac {m}{2\pi} \left( -\pi +\frac{2\arctan (\Gamma/|m|)}{|m|} -\frac{2\Gamma m}{\Gamma^2 +m^2}\right) .\label{k00}
\end{equation}
In the PV1 scheme, one has
\begin{equation}
    \sigma_{\mathrm{PV1}}(0,m,\Gamma)=\frac {m}{\pi} \left( \frac{\arctan (\Gamma/|m|)}{|m|} -\frac{\Gamma m}{\Gamma^2 +m^2}\right).\label{k00PV1}
\end{equation}
In the PV2 scheme, the conductivity vanishes identically in this limit, 
$\sigma_{\mathrm{PV2}}(0,m,\Gamma)=0$
which also happens at $\Gamma=0$ in any scheme in the absence of a chemical potential. (Note, that the limits $k\to 0$ and $m\to 0$ do not commute.) Although there are analytic formulas (\ref{k00}) and (\ref{k00PV1}) we present the plots on Fig.\ \ref{fig:k00}) for convenience.

Essential differences between two schemes appear, as expected, when $\Gamma$ is large. In the infinite $\Gamma$ limit,
\begin{equation}
    \lim_{\Gamma\to\infty}\sigma_{\mathrm{PV2}}=0\label{PV2Ginfty}
\end{equation}
in contrast to (\ref{PV1Gammatoinfty}). Already at a finite $\Gamma$ the differences are significant. For a finite $\Gamma$ and $m=0$, one has
\begin{equation}
    \sigma_{\mathrm{PV2}}(k,0,\Gamma)=0,\label{PV2m0}
\end{equation}
while in PV1 this value is nonzero, see (\ref{PV1m0}). 
\begin{figure}[h]
    \centering
    \includegraphics[width=16cm]{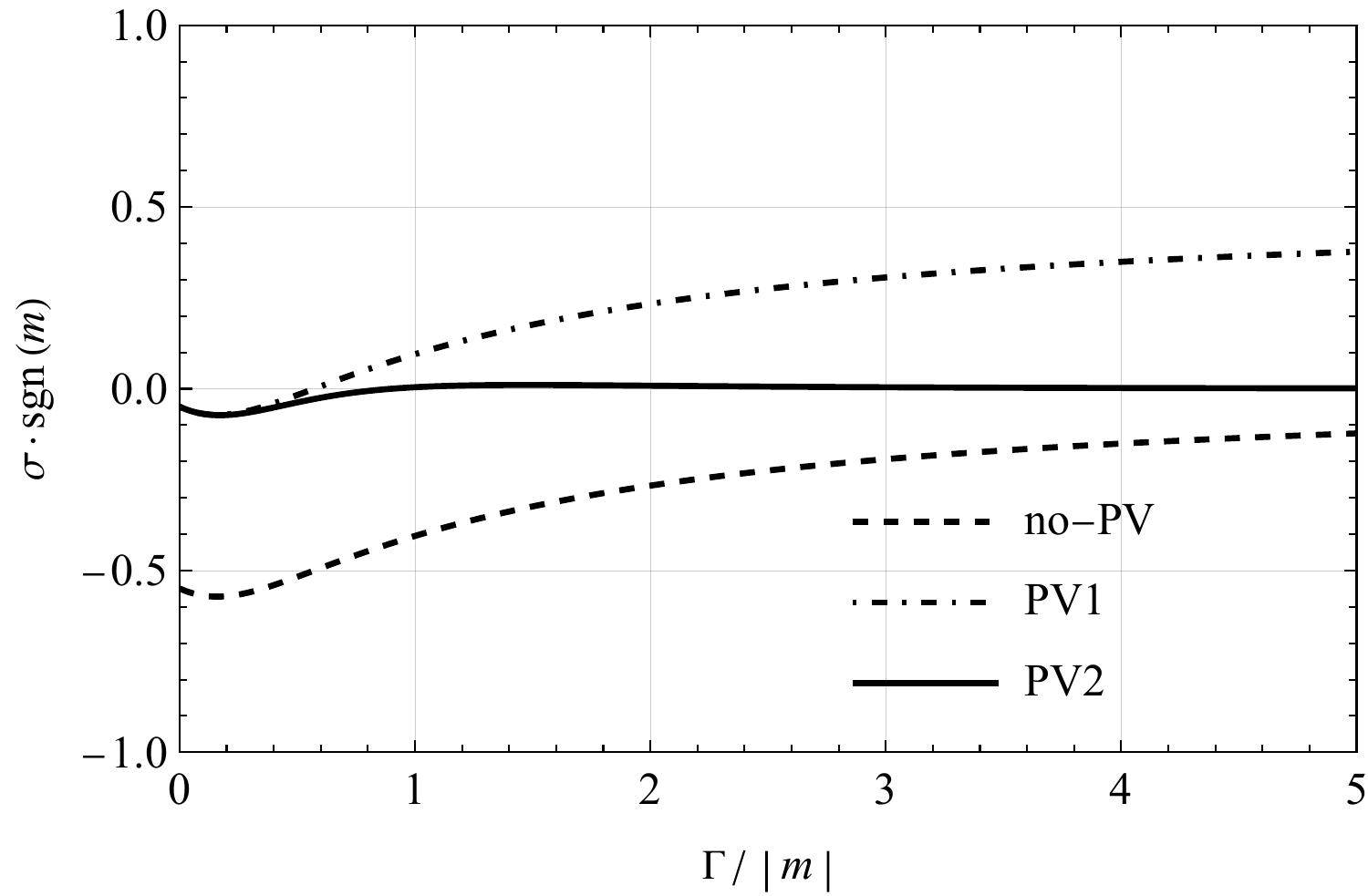}
    \caption{The anomalous conductivity $\sigma$ as a function of $\Gamma/|m$ for $\Vec{k}=0$, $k_0/|m|=1 $. The dashed line corresponds to the conductivity without subtraction, while the dashed-dotted and solid lines correspond to PV1 and PV2 schemes, respectively.}
    \label{fig:k0eq05}
\end{figure}
\begin{figure}[h]
\centering
\includegraphics[width=16cm]{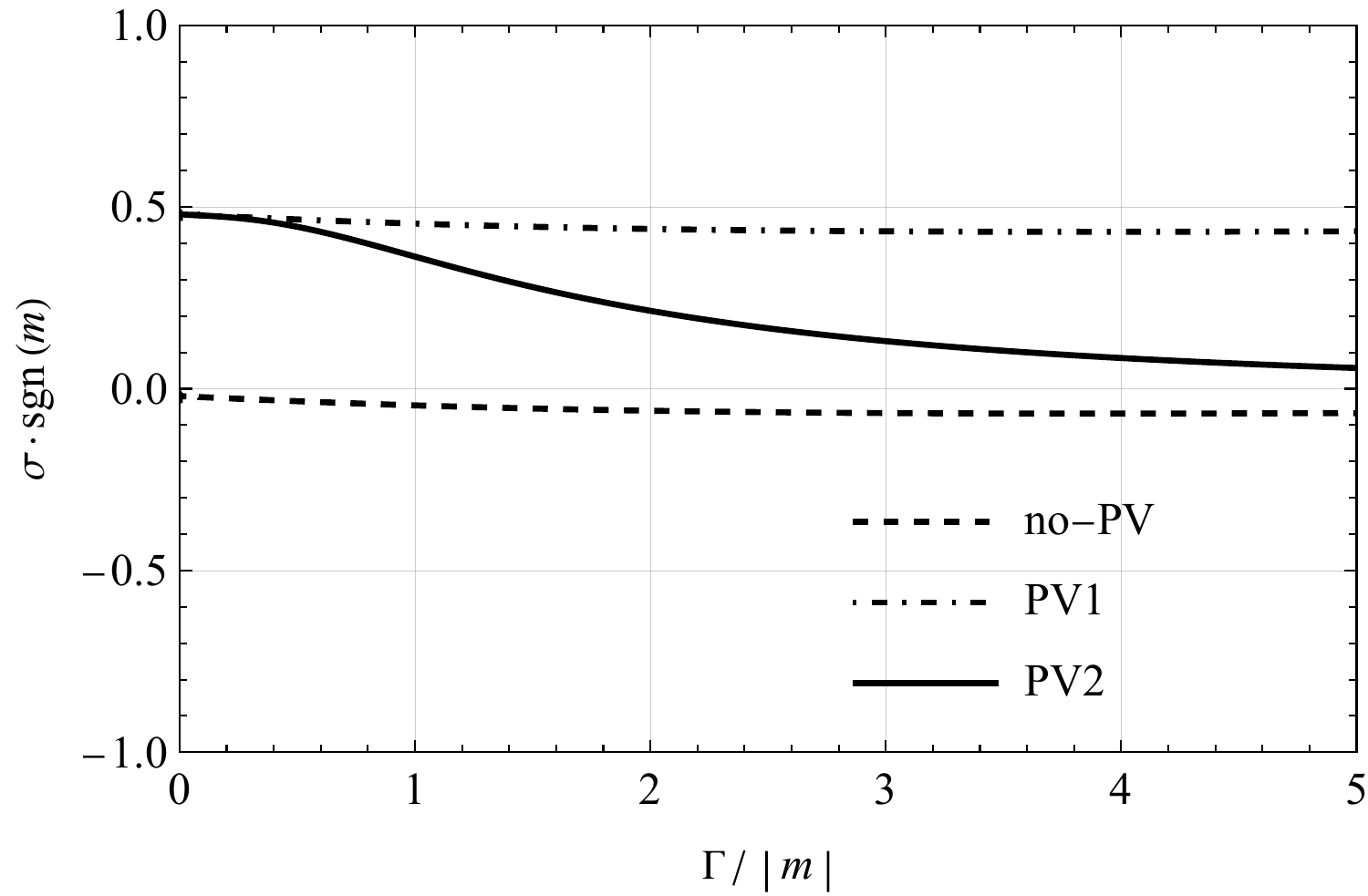}
\caption{The anomalous conductivity $\sigma$ as a function of $\Gamma/|m$ for $\Vec{k}=0$, $k_0/|m|=10 $. The dashed line corresponds to the conductivity without subtraction, while the dashed-dotted and solid lines correspond to PV1 and PV2 schemes, respectively.}
\label{fig:k0eq5}
\end{figure}

For $k_0/|m|=1$ and $k_0/|m|=10$ the conductivity $\sigma$ is depicted at Fig.\ \ref{fig:k0eq05} and Fig.\ \ref{fig:k0eq5}, respectively. In both cases $\Vec{k}=0$. Note, that large values of $\Gamma/|m|$ do not necessarily mean that impurities are strong. Equally, the mass gap may be small. In general, in the scheme PV2, the impurities damp anomalous Hall conductivity, while in PV1 they do not. For a moderate frequency, see Fig.\ \ref{fig:k0eq05}, $\sigma_{\mathrm{PV2}}$ is close to zero for all values of the parameter $\Gamma$.

In $2+1$ dimensions the structure of ultraviolet divergences is quite simple so that just a single regulator field is required. The only essential feature is that one has to take the limit of an infinite mass gap of the regulator field. From this point of view, there are good reasons to expect that both subtraction schemes are internally consistent and remove all ultraviolet divergences. These arguments are however not watertight and we shall pay more attention to the QFT aspects of Pauli--Villars scheme in some future work. Note, that there are examples when modifications of a model impose severe restrictions on the PV subtractions. (See e.g. the paper \cite{Fialkovsky:2019rum} where the PV scheme in 4D QED with boundaries was analyzed.)

At any rate, the final word in choosing between two subtraction schemes should belong to an experiment.





\section{Discussion}\label{sec:dis}
In this paper, we have studied with Quantum Field Theory Methods the anomalous part of polarization tensor of a $2+1$ dimensional fermion interacting with impurities described by a scattering rate $\Gamma$ for arbitrary external frequency and momenta. We used the PV scheme and argued that there are two natural subtractions. In one of them, which we called PV1, the contribution of a massive regulator field is subtracted in the limit $|M|\to\infty$. This is just the standard PV subtraction known from text books. In the other scheme, we treat $\Gamma/m$ as a dimensional parameter similar to the electric charge. With such an interpretation, it is natural to assume that the regulator field sees the same ``impurity charge" $\Gamma_R/M=\Gamma/m$. This boils down to subtracting a double limit $|M|,\Gamma_R\to\infty$ while keeping the ratio $\Gamma_R/M$ fixed. In this scheme, called PV2 throughout this work, the dependence of Hall conductivity $\sigma$ on $\Gamma$ differs crucially from the predictions of PV1. In PV1, the conductivity is enhanced by the impurities while in PV2 the impuirities tend to diminish $\sigma$. The last word in choosing between PV1 and PV2 should belongs to an experiment. However, further consistency checks are also needed. An important lesson from the papers \cite{Deser:1997gp,Deser:1997nv} is that large gauge invariance cannot be used to check perturbative calculations since corresponding Ward identities contain important nonperturbative contributions.

There are alternative methods of calculations of the parity anomaly. One of them relies on the $\zeta$ function regularization. It allows to evaluate the anomaly for massless \cite{AlvarezGaume:1984nf} and massive \cite{Deser:1997gp,Deser:1997nv,Fosco:1997ei} fermions, in the presence of an external magnetic field \cite{Beneventano:2009uv}, and even in the presence of boundaries  \cite{Kurkov:2017cdz}. The results obtained with this method are consistent with the Pauli--Villars scheme. Unfortunately, there is no generalization of the $\zeta$ function regularization in the presence of impurities.
For completeness, we like to mention a rather extreme  proposal \cite{DelCima:2009sm} that the parity anomaly is merely a counterterm which is needed to restore parity.

The Casimir force between surfaces which exhibit a Hall-type conductivity may become repulsive (for an overview of this effect see \cite{Fialkovsky:2018fpo,Lu:2021jvu} and a recent paper \cite{Canfora:2022xcx}). The Casimir interaction is an integral effect. That is, all frequencies and momenta contribute to the force. The study of this effect was one of the main motivations for the calculation reported above. At present, we may suggest that in the repulsion will be most probably damped by impurities in the PV2 scheme and enhanced in PV1.

\begin{acknowledgments} 
One of us (D.V.) is grateful to Ignat Fialkovsky for previous collaboration and discussions on impurities. The work of D.V. was supported in parts by the S\~ao Paulo Research Foundation (FAPESP), grant 2021/10128-0, and by the National Council for Scientific and Technological Development (CNPq), grant 304758/2022-1. O.H. acknowledges support by the S\~ao Paulo Research Foundation (FAPESP), by the grant
2019/26291-8. The work of R.~M. was 
funded by the Deutsche Forschungsgemeinschaft (DFG, German Research Foundation)
through Project-ID 258499086—SFB 1170 ToCoTronics and through the Würzburg-Dresden Cluster 
of Excellence on Complexity and Topology in Quantum Matter – ct.qmat Project-ID 
390858490—EXC 2147.
\end{acknowledgments}

\bibliography{export}

\end{document}